\documentstyle[11pt,epsfig]{article}
\newcommand{\lwig}{\mbox{\,\raisebox{.3ex}
{$<$}$\!\!\!\!\!$\raisebox{-.9ex}{$\sim$}\,}}

\begin{document}
\date{}

\title{{\normalsize\rightline{DESY 95-119}\rightline{hep-ph/9506392}}
        \vskip 1cm
        \bf QCD--Instanton Induced Final States in \\
        \vglue 0.2cm
        Deep Inelastic Scattering\thanks{Talk given in the WG II,
        Session C, at the Workshop on Deep Inelastic Scattering and
        QCD, Paris, April 1995}\\
        \vspace{11mm}}

\author{M. Gibbs$^a$, A. Ringwald$^b$ and F.
Schrempp$^b$\\[0.5cm]
 $^a$Oliver Lodge Laboratory, University of Liverpool, Liverpool, UK\\
 $^b$Deutsches Elektronen-Synchrotron DESY, Hamburg, Germany\\ \vspace{2mm}}
\begin{titlepage}
\maketitle
\begin{abstract}
We report briefly on a
broad and systematic study of possible manifestations
of QCD-instantons at HERA. We concentrate on the high multiplicity
final state structure, reminiscent of an isotropically decaying
``fireball''. First results of a Monte Carlo simulation are presented,
with emphasis on the typical event-structure and the  transverse
energy, muon and $K^0$ flows.
\end{abstract}

\thispagestyle{empty}
\end{titlepage}
\newpage
\setcounter{page}{2}

\section{Introduction}
The Standard Model of electro-weak (QFD) and strong (QCD) interactions
is remarkably successfull. Its perturbative formulation (``Feynman
diagrammatics'') appears to be theoretically consistent and agrees with
present experiments.
Nevertheless, even for small couplings, there exist processes that
cannot be described by  conventional perturbation theory and, notably,
violate the classical conservation laws of certain fermionic quantum
numbers~\cite{th} (see fig.\,1).

%%%%%%%%%%%%%%%%%%%%%FIGURE 1%%%%%%%%%%%%%%%%%%%%%%%%%%%%%%%%%
\begin{figure} [bht]
\vspace{-0.4cm}
\begin{center}
\epsfig{file=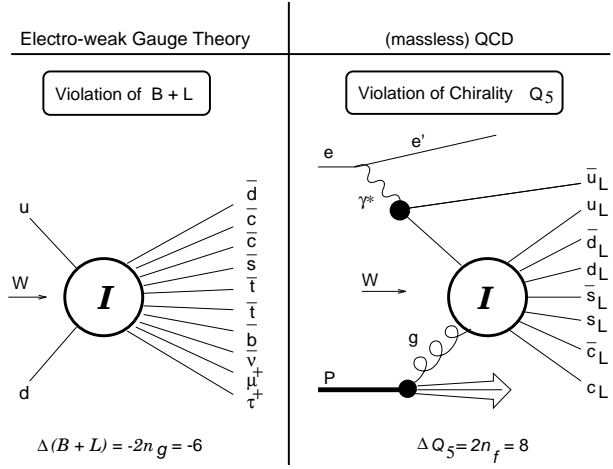,angle=-90,width=8cm}
\caption[dum]{The basic anomalous processes induced
by instantons in QFD and QCD, respectively}
\end{center}
\end{figure}
%%%%%%%%%%%%%%%%%%%%%%%%%%%%%%%%%%%%%%%%%%%%%%%%%%%%%%%%%%%%%

Such anomalous processes are induced by {\it
instantons}~\cite{bpst} which represent tunnelling processes in Yang-Mills
gauge theories, associated with the highly degenerate
vacuum structure.

For many years, such tunnelling transitions have been considered
largely of academic interest, due to their exponential suppression
$\propto \exp{(-4\pi/\alpha)}$ at low energies. A few years ago, however,
much activity in this field was generated by the observation~\cite{r}
that instanton-induced processes may well become unsuppressed,
i.\,e. observable, {\it at high energies}.

The basic significance and possible importance of QCD-instanton
effects in deep inelastic scattering (DIS) for decreasing  Bjorken variable
$x_{\rm Bj}$ and high photon virtuality $Q^2$
has recently been emphasized~\cite{bb}:
\begin{itemize}
\item First of all, QCD-instanton effects for decreasing $x_{\rm Bj}$ are
largely analogous~\cite{kr} to the manifestation
of electro-weak  instantons at increasing energies.
The anomalous $B+L$ violation due to electro-weak instantons
is paralleled by a chirality violation induced by
QCD-instantons~\cite{th} (c.\,f. fig.\,1).

\item Secondly, discovery of QCD-instanton induced DIS-events would
itself be of basic significance, since they correspond to a novel,
non-perturbative manifestation of QCD.
\end{itemize}

Whereas a promising search for anomalous
electro-weak events is only possible in the far future, presumably
at a post-LHC collider~\cite{ewcollpheno} or at
cosmic ray facilities~\cite{ewcosmray}, the search for anomalous events induced
by QCD-instantons can start right now,
in deep inelastic $e^\pm p$ scattering at HERA.

%%%%%%%%%%%%%%%%%%%%%FIGURE 2%%%%%%%%%%%%%%%%%%%%%%%%%%%%%%%%%
\begin{figure} [hbt]
\vspace{-0.4cm}
\begin{center}
\epsfig{file=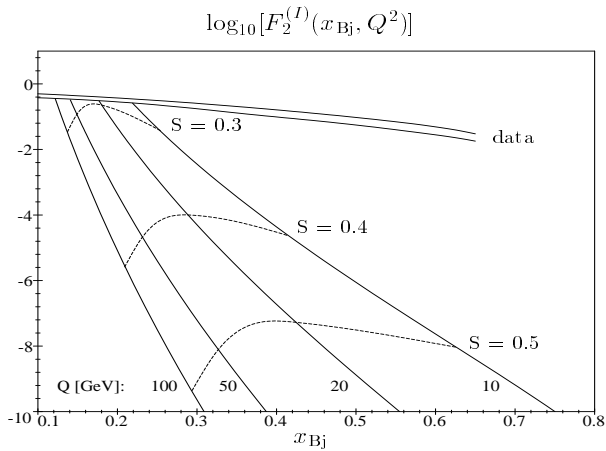,bbllx=113pt,bblly=277pt,%
bburx=491pt,bbury=617pt,width=8cm,height=6cm}
\caption[dum]{The logarithm of the instanton-induced contribution to
the structure function  $F_2$ of the proton.
The curves denoted by ``data'' roughly represent
the trend of the experimental data for $F_2$.}
\end{center}
\end{figure}
%%%%%%%%%%%%%%%%%%%%%%%%%%%%%%%%%%%%%%%%%%%%%%%%%%%%%%%%%%%%%
Naturally, the first observables where manifestations of
QCD-instantons may be looked for are the nucleon structure functions.

Within the theoretical framework of Ref.~\cite{bb},
we have performed~\cite{rs}
a state of the art evaluation of the instanton-induced
contribution to $F_2(x_{\rm Bj},Q^2)$
(see fig.\,2).
It rises strongly with decreasing $x_{\rm Bj}$ and tends
to reach the size of the experimental data around
$x_{\rm Bj} \approx 0.1 \div 0.25$. Unfortunately, due to
inherent uncertainties, the calculation cannot be trusted
anymore for $x_{\rm Bj}\lwig 0.35$, say.
Nevertheless, the trend is very suggestive!

There are a number of reasons~\cite{rs} that favour  experimental
searches for instanton-induced ``footprints'' in the
multi-particle final state over searches via the
structure functions, the latter being the most inclusive observables in
deep
inelastic scattering. It is the purpose of this contribution to present
a brief status report on our phenomenological analysis of the instanton
induced final state.
\section{The Instanton Induced Final State}

\subsection{Characteristic Features}

The instanton-induced contribution to the multi-particle final state in
DIS arises in form of an instanton-induced subprocess (denoted by ``$I$''
in fig.\,3) along with a current-quark jet. The relevant kinematical
variables are summarized in table\ 1.

%
%%%%%%%%%%%%%%%%%%%%%FIGURE 3%%%%%%%%%%%%%%%%%%%%%%%%%%%%%%%%%
\begin{figure} [bht]
\vspace{-0.4cm}
\begin{center}
\epsfig{file=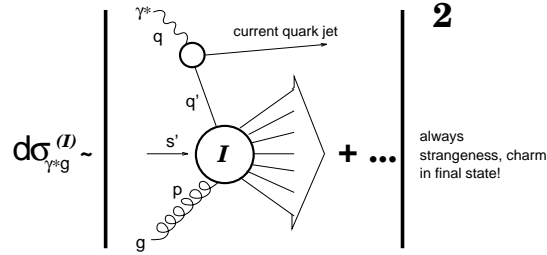,angle=-90,width=7cm}
\caption[dum]{Structure and kinematics of the instanton-induced contribution to
the $\gamma^\ast g$ cross section}
\end{center}
\end{figure}
%%%%%%%%%%%%%%%%%%%%%%%%%%%%%%%%%%%%%%%%%%%%%%%%%%%%%%%%%%%%%
%
\begin{table} [hbt]
\begin{center}
\begin{tabular}{c|c|c}   &
$\gamma^\ast g$ & $I$-subprocess  \\ \hline &
\rule[0mm]{0mm}{4mm}
$Q^2=-q^2$ & $Q^{\prime 2}=-q^{\prime 2}$  \\
\raisebox{1.5ex}[-1.5ex]{Bj.-Variables} &
$x={Q^2\over 2pq}$ & $x^\prime={Q^{\prime 2}\over 2pq^\prime}$  \\
\end{tabular}
\end{center}
\caption{Relevant kinematical variables, with the primed quantities
referring to the $I$-subprocess in fig.\,3.
Note that $x_{\rm Bj}<x<x^\prime<1$.}
\end{table}
%%%%%%%%%%%%%%%%%%%%%FIGURE 4%%%%%%%%%%%%%%%%%%%%%%%%%%%%%%%%%
\begin{figure}
\vspace{-0.4cm}
\begin{center}
\epsfig{file=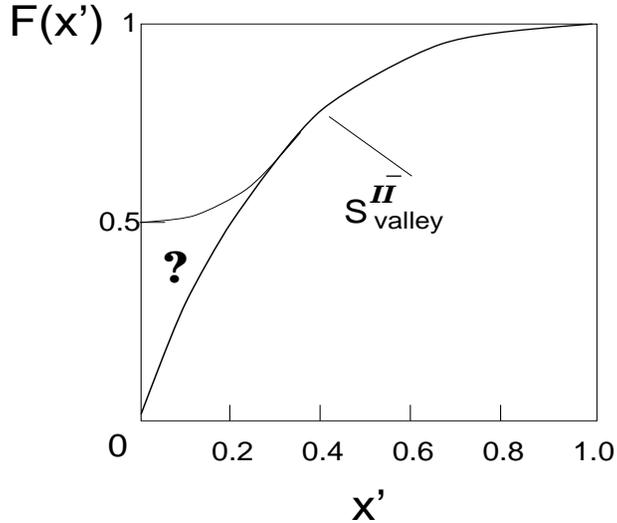,angle=-90,width=8cm,height=7cm}
\caption[dum]{``Holy grail'' function $F(x^\prime)$.
The lower curve corresponds to the prediction from the $I\bar{I}$
valley method, with $S^{I\overline{I}}_{\rm valley}$ denoting the
valley action.}
\end{center}
\end{figure}
%%%%%%%%%%%%%%%%%%%%%%%%%%%%%%%%%%%%%%%%%%%%%%%%%%%%%%%%%%%%%
Our phenomenological analysis is based on the following
set of characteristic features which have emerged from various
theoretical investigations:

\begin{itemize}
  \item {\it Isotropic} emission of many semi-hard
         partons in the $I$-rest system,
         $\vec{q^\prime}+\vec{p}=0$,
         reminiscent of a ``decaying fireball''.
  \item {\it High multiplicity}: $\langle n_{q+g}^{(I)}\rangle_{n_f=4}
         \simeq 10$ (!) at HERA~\cite{rs}.
  \item Characteristic flavour flow ({strangeness, charm})~\cite{th}.
  \item {Strong peaking} of the $I$-subprocess total cross section~\cite{bb,kr}
         for decreasing Bjorken variables { $x^\prime, Q^{\prime 2}$}:
\begin{eqnarray}
\nonumber
 s^\prime \sigma_{\rm tot}^{(I)} (x^\prime ,Q^{\prime 2}) \sim
\exp\Biggl\{ -\frac{4\pi}{\alpha_s^{\rm eff}(Q^{\prime 2})}F(x^\prime )\Biggr\}
\end{eqnarray}
The importance of instanton-induced events at small $x^\prime$ crucially
relies on the precise functional dependence of the ``holy grail''
function $F(x^\prime)$ (c.\,f. fig.\,4). Unfortunately, it is reliably
known only at large $x^\prime\sim 1$ within instanton-perturbation
theory. By means of the $I\bar{I}$ valley method~\cite{kr}
$F(x^\prime)$ is obtained beyond perturbation theory for {\it all}
$x^\prime$, however with intrinsic ambiguities as typically depicted in
fig.\,4.
\end{itemize}

{}From the above  properties, the qualitative event structure is expected to
consist of a current-quark jet along with a densely populated
hadronic ``band''   in the $(\eta_{\rm lab},\phi_{\rm lab})$-plane~\cite{rs}.
The hadronic band directly reflects the isotropy in the $I$-rest
system along with a large multiplicity of semi-hard quarks and gluons.

\subsection{Monte Carlo Simulation}

A Monte Carlo simulation for instanton-induced events at HERA
(QCD INSTANTON MC 1.0), based on HERWIG 5.8, has been essentially
completed~\cite{grs}. It includes full hadronization.
While it should already account for the composition of the
final state quite well, we do not yet quote absolute production rates,
since a {\it reliable} theoretical estimate of the cross sections, in
particular at smaller values of $x^\prime$ and $Q^{\prime 2}$, is
quite difficult and still in progress~\cite{rsth,mrs}.

The main questions addressed so far, concern the typical
event-structure, and transverse energy as well as muon and $K^0$ flows
and some characteristic event distributions.

Let us present the results for a sample of instanton-induced events,
corresponding to an intermediate small $x^\prime$ behaviour of
the ``holy grail'' function $F(x^\prime)$ (c.\,f. fig.\,4):

$$
F(x^\prime)=\left\{
\begin{array}{ll}
S^{I\overline{I}}_{\rm valley}(x^\prime ),\   &{\rm for}\
x^\prime\geq 0.12 , \\
{\rm const.},\ &{\rm for}\        x^\prime < 0.12
\end{array}
\right.
$$
Throughout, we take $x_{\rm Bj}\geq 10^{-3}, y_{\rm Bj}\geq 0.1$.
Furthermore, we impose a cut on the total invariant mass in the $I$-subprocess,
$\sqrt{s^\prime}\ge 10$ GeV, in order to guarantee a minimum virtuality
$Q^{\prime 2}\ge 1$ GeV$^2$.

In fig.\,5 a typical instanton-induced event is displayed. The
current-quark jet (around $\eta_{\rm lab}\simeq -0.5$)
along with the expected,
densely populated  hadronic ``band'', centered around $\eta_{\rm
lab}\simeq 2.5$, are apparent. The electron is shown here as well.

In fig.\,6, the transverse energy flow versus
$\eta_{\rm lab}$ is displayed. It  exhibits a strong enhancement across the
hadronic ``band'', since each of the many hadrons from the $I$-subprocess
contributes a comparable energy into a single $\eta_{\rm lab}$ bin of
width $\approx 1.8$.

A related important quantity is the distribution of the transverse energy
for hadrons within the ``band'' (fig.\,7). It peaks around $E_t\simeq
\sqrt{s^\prime}_{\rm min}=10$ GeV, i.\,e. at a value much larger than
the one from present experimental data.

The $K^0$ and muon flows, displayed in figs.\,8 and 9, again peak
at the center of the band of hadrons emerging from the $I$-subprocess.
This is presumably a distinctive signature for instanton
induced events. It directly reflects
the basic fact that in {\it each} such event a pair of
strange and of charmed quarks is produced (c.\,f. fig.\,1).
We find $\langle N_{K^0_S}\rangle\simeq 1.8$ and  $\langle
N_{\mu}\rangle\simeq 0.2$.

\section{Conclusion}

A systematic phenomeological and theoretical
investigation of the discovery potential for QCD-instanton induced
%%%%%%%%%%%%%%%%%%%%%FIGURE 5%%%%%%%%%%%%%%%%%%%%%%%%%%%%%%%%%
\begin{figure} [hbt]
\vspace{1.4cm}
\begin{center}
\epsfig{file=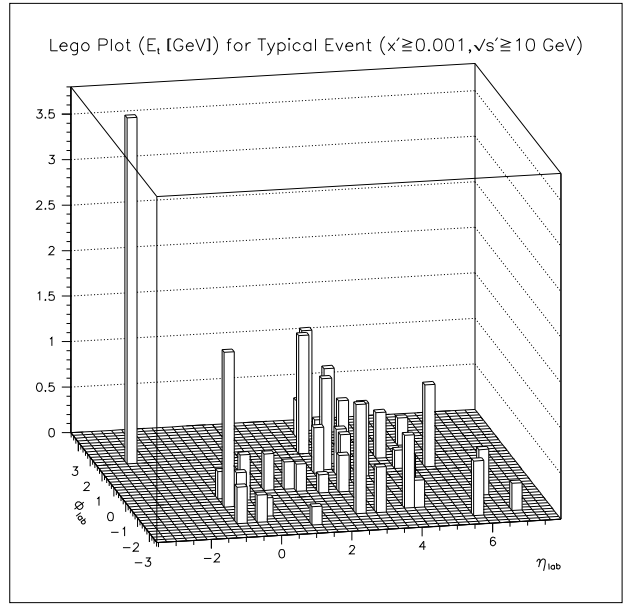,bbllx=19pt,bblly=149pt,bburx=531pt%
,bbury=672pt,width=7.5cm,height=7.5cm}
\caption[dum]{ A typical instanton-induced event in the lab. system}
\end{center}
\end{figure}
%%%%%%%%%%%%%%%%%%%%%%%%%%%%%%%%%%%%%%%%%%%%%%%%%%%%%%%%%%%%%
%%%%%%%%%%%%%%%%%%%%%FIGURE 6%%%%%%%%%%%%%%%%%%%%%%%%%%%%%%%%%
\begin{figure}
\vspace{-0.4cm}
\begin{center}
\epsfig{file=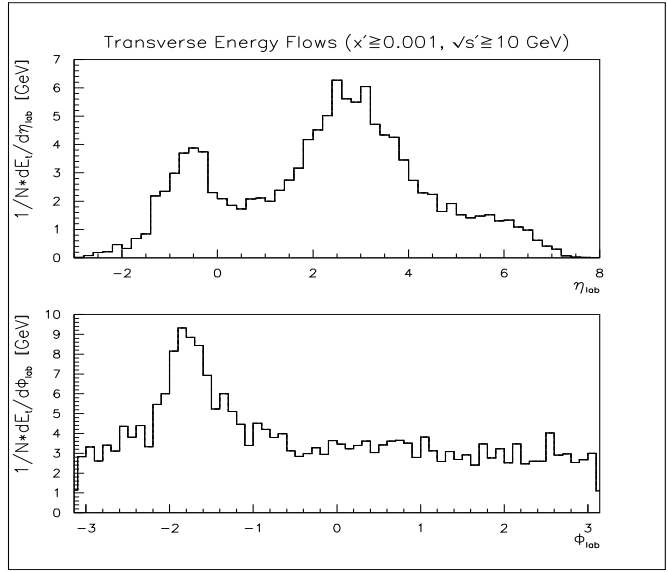,bbllx=22pt,bblly=156pt,bburx=531pt%
,bbury=672pt,width=8cm,height=7cm}
\caption[dum]{ Hadronic transverse energy flows versus $\eta_{\rm lab}$ and
$\phi_{\rm lab}$, respectively}
\end{center}
\end{figure}
%%%%%%%%%%%%%%%%%%%%%%%%%%%%%%%%%%%%%%%%%%%%%%%%%%%%%%%%%%%%%
%%%%%%%%%%%%%%%%%%%%%FIGURE 7%%%%%%%%%%%%%%%%%%%%%%%%%%%%%%%%%
\begin{figure}
\vspace{-0.4cm}
\begin{center}
\epsfig{file=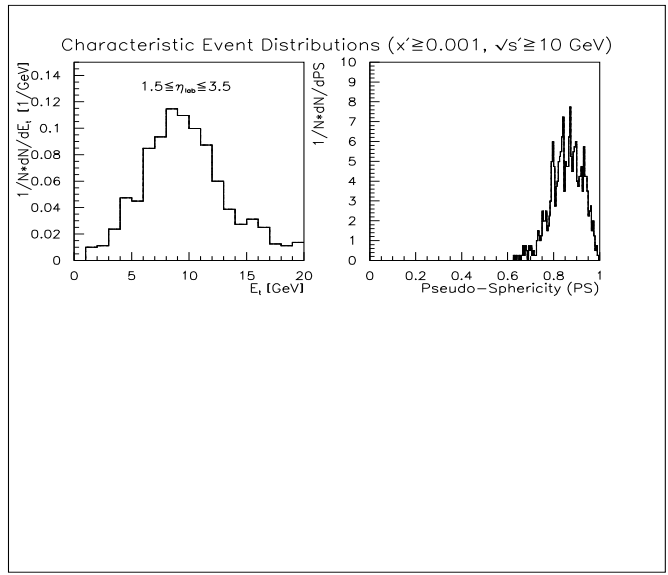,bbllx=22pt,bblly=156pt,bburx=531pt%
,bbury=672pt,width=8cm,height=7cm}
\caption[dum]{ Transverse energy and pseudo-sphericity distributions}
\end{center}
\end{figure}
%%%%%%%%%%%%%%%%%%%%%%%%%%%%%%%%%%%%%%%%%%%%%%%%%%%%%%%%%%%%%
%%%%%%%%%%%%%%%%%%%%%FIGURE 8%%%%%%%%%%%%%%%%%%%%%%%%%%%%%%%%%
\begin{figure}
\vspace{-0.4cm}
\begin{center}
\epsfig{file=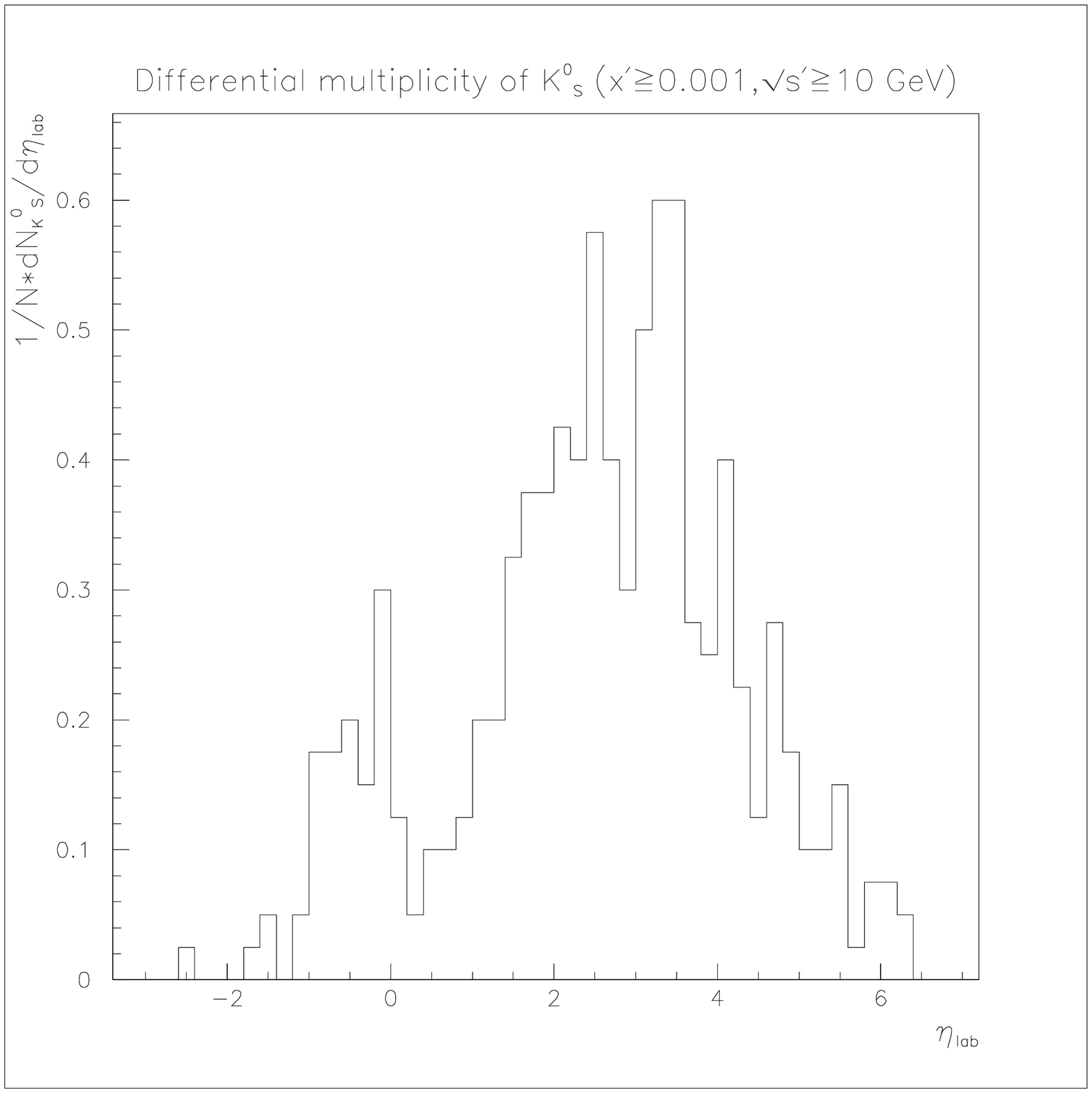,bbllx=22pt,bblly=156pt,bburx=531pt,bbury=672pt%
,width=8cm,height=7cm}
\caption[dum]{Flow of $K^0_S$ versus $\eta_{\rm lab}$, peaking around the
center of the hadronic ``band'' from the $I$-subprocess}
\end{center}
\end{figure}
%%%%%%%%%%%%%%%%%%%%%%%%%%%%%%%%%%%%%%%%%%%%%%%%%%%%%%%%%%%%%
%%%%%%%%%%%%%%%%%%%%%FIGURE 9%%%%%%%%%%%%%%%%%%%%%%%%%%%%%%%%%
\begin{figure}
\vspace{-0.4cm}
\begin{center}
\epsfig{file=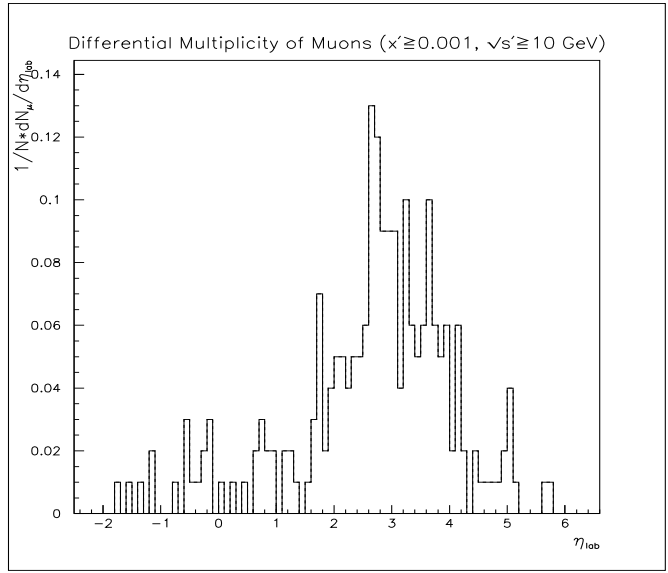,bbllx=22pt,bblly=156pt,bburx=531pt,bbury=672pt%
,width=8cm,height=7cm}
\caption[dum]{ Flow of muons versus $\eta_{\rm lab}$, peaking around the
center of the hadronic ``band'' from the $I$-subprocess}
\end{center}
\end{figure}
%%%%%%%%%%%%%%%%%%%%%%%%%%%%%%%%%%%%%%%%%%%%%%%%%%%%%%%%%%%%%
events is under way. Clearly, HERA offers a unique window in DIS!
A discovery of such events would be of basic importance: first of all,
as a novel, non-perturbative manifestation of QCD and secondly,
because of the close analogy to anomalous $B+L$ violation in
electro-weak processes in the multi-TeV regime.

\section*{Acknowledgements}

It is a pleasure to thank F. Botterweck, M. Kuhlen
and A. de Roeck for helpful discussions on experimental aspects
and suggestions.
\vspace{0.2cm}

\end{document}